\documentclass[twocolumn,aps,prc,showpacs,preprintnumbers]{revtex4}%
\usepackage{amssymb}
\usepackage{amsmath}
\usepackage{amsfonts}
\usepackage{graphicx}
\usepackage{dcolumn}
\usepackage{bm}%
\providecommand{\U}[1]{\protect\rule{.1in}{.1in}}
\providecommand{\U}[1]{\protect\rule{.1in}{.1in}} \textheight 9.7in
\begin{document}
\title{Electron screening and its effects on Big-Bang nucleosynthesis}
\author{ Biao~Wang$^{(1)}$, C.A.~Bertulani$^{(1)}$,and A.~B.~Balantekin$^{(2)}$}
\email{biao_wang@leo.tamu-commerce.edu, carlos_bertulani@tamu-commerce.edu, 
 baha@physics.wisc.edu}
\address{$^{(1)}$Department of Physics, Texas A\&M University, Commerce,
TX 75429, USA\\
$^{(2)}$Physics Department, University of Wisconsin-Madison, 
1150 University Avenue, Madison, WI 53706. }

\begin{abstract}
We study the effects of electron screening on nuclear reaction rates occurring during the Big Bang nucleosynthesis epoch. The sensitivity of the predicted elemental abundances on electron screening  is studied in details. It is shown that electron screening does not produce noticeable results in the abundances unless the traditional Debye-H\"uckel model for the treatment of electron screening in stellar environments is enhanced by  several orders of magnitude.  The present work rules out electron screening as a relevant ingredient to Big Bang nucleosynthesis confirming a previous study by Itoh {\it et al.} \cite{Itoh97} and ruling out exotic possibilities for the treatment of screening, beyond the mean-field theoretical approach.

\end{abstract}
\date{\today}

 \keywords{}\maketitle

During the Big Bang the Universe evolved very rapidly and only the lightest nuclides (e.g., D, $^3$He, $^4$He, and
$^7$Li) could be synthesized. The abundances of these nuclides  are probes of the conditions of the Universe during the
very early stages of its evolution, i.e., the first few minutes. The conditions during the Big Bang nucleosynthesis (BBN) are believed to be well described in terms of standard models of cosmology and particle physics which determine the values of, e.g., temperature, nucleon density, expansion rate, neutrino content, neutrino-antineutrino asymmetry, etc. Deviations from the BBN
test the parameters of these models, and constrain nonstandard physics or cosmology that may alter the conditions during BBN \cite{WFH67,Iocco:2008va}. Sensitivity to the several parameters and physics input in the BBN model have been investigated thoroughly in the past (see, e.g., \cite{Nol,Serpico:2004gx,Cyburt:2001,Cyburt:2002,Cyburt:2004cq,Ful10}).  In this work we consider if the screening by electrons would have any impact on the BBN predictions. This work reinforces the conclusions presented in Ref. \cite{Itoh97}, namely, that screening is not a relevant ingredient of the BBN nucleosynthesis.

Modeling the BBN and  stellar evolution  requires that one includes  the information on nuclear reaction rates $\langle \sigma v\rangle$ in reaction network calculations, where $\sigma$ is the nuclear fusion cross sections and $v$ is the relative velocity between the participant nuclides. Whereas $v$ is well described by a Maxwell-Boltzmann velocity distribution for a given temperature $T$, the cross section $\sigma$ is taken from laboratory experiments on earth, some of which are not as well known as desired \cite{Nol,Serpico:2004gx,Cyburt:2001,Cyburt:2002,Cyburt:2004cq}. The presence of atomic electrons in laboratory experiments also influence the measured values of the cross sections in a rather unexpected way (see, e.g. \cite{BBH97}). In the network calculations for the description of elemental synthesis in the BBN or in stellar evolution one needs account for the differences of the ``bare" nuclear cross sections obtained in laboratory measurements, \textit{$\sigma_b$}$(E)$, to the corresponding quantities in stellar interiors, \textit{$\sigma_s$}$(E)$. One of these corrections is due to stellar electron screening as light nuclei in the stellar environments are almost completely ionized. 

Using the Debye-H\"uckel model, Salpeter \cite{Sal54} showed that stellar electron screening enhances cross sections,
reducing the Coulomb barrier that reacting ions must overcome, yielding an enhancement factor 
\begin{equation}
 f(E) = {\sigma _{s}(E) \over \sigma _{b}(E)}.
    \label{f0}
\end{equation}
The  Debye-Hueckel model used by Salpeter yields a screened Coulomb potential, valid when $\langle V\rangle \ll kT$ (weak screening),
\begin{equation}
V(r) = {e^2 Z_i \over r} \exp\left(-{r \over R_D} \right),
\end{equation}
which depends on the ratio of the Coulomb potential
at the Debye radius $R_D$ to the temperature,
\begin{equation}
    f =\;\mathrm{exp} \left({Z_1 \,Z_2 \;e^2 \over R_D k T}\right)
    =\;\mathrm{exp} \left( {0.188\,Z_1 \,Z_2 \;\zeta \;\rho
    ^{1/2}\;T_6^{-3/2} } \right),
        \label{f02}
\end{equation}
where 
\begin{equation}
\zeta R_D = \left( {k T \over 4 \pi e^2 n} \right)^{1/2},
\label{DR}
\end{equation}
is the Debye radius,
$n$ is the ion number density, $\rho$ is a dimensionless quantity
measured in units of g/cm$^3$,
\begin{equation}\zeta = \left[ \sum\limits_i~X_i ~
\frac{Z_i^2}{A_i} +  \chi \sum\limits_i~X_i \frac{Z_i}{A_i} \right]^{1/2},\end{equation}
where $X_i$ is the mass fraction of nuclei of type $i$,
and $T_6$ is the dimensionless temperature in units of 10$^6$ K. The factor
$\chi$ corrects for the effects of electron degeneracy \cite{Sal54}.

Corrections 
to the Salpeter formula are expected at some level.  
Nonadiabatic effects have been suggested as one source, e.g.,  
a high Gamow energy allows reacting nuclei having velocities significantly higher than the typical
ion velocity, so that the response of slower plasma ions might be suppressed.   Dynamic corrections were first discussed 
by \cite{Mit77} and later studied by \cite{CSK88}.
Subsequent work showed that Salpeter's formula would be valid
independent of the Gamow energy due to the nearly precise thermodynamic
equilibrium of the solar plasma \cite{BS97,Gru98,GB98}. 
Later, a number of contradictions were pointed out in
investigations claiming larger corrections, and a field theoretic aproach
was shown to  lead to the expectation
of only small ($\sim$ 4\%) corrections to the standard formula, for solar conditions \cite{BBGS02}. 

Controversies about the magnitude of the screening effect have not entirely died out and continued in some works \cite{Sha96,Sha00,Mao09,WFT01}. These works are invariably based  on molecular dynamics simulations. Dynamic screening becomes important because the
nuclei in a plasma are much slower than the electrons and are not able to rearrange themselves as quickly around faster moving ions.
Since nuclear reactions require energies several times the average thermal energy, the ions that are able to engage in nuclear reactions in the stars are such faster moving ions, which therefore may not be accompanied by their full screening cloud. In fact, dynamic effects
are important when particles react with large relative velocities.  It appears that pairs of ions with greater relative velocities
experience less screening than pairs of particles with lower relative velocities \cite{Mao09}. The correction for dynamical screening can be approximated by replacing $R_D$ in eq. \eqref{f02} by a velocity dependent quantity $R_D(v_p)=R_D\sqrt{1+\mu v_p^2/kT}$, where $\mu$ is the ion pair reduced mass and $v_p$ their relative velocity \cite{KSK05}.
This effect is relevant in stellar environments whenever nuclear reactions  occur at energies that are greater than the thermal energy.

Experimentalists have exploited surrogate environments to test our understanding of plasma screening effects.
For example, screening in d(d,p)t has been studied for gaseous targets and for deuterated metals, insulators, and semiconductors \cite{Rai04,HPR05}.   It is believed that the quasi-free valence
electrons in metals create a screening environment quite similar to that found in stellar plasmas.  The experiments in metals seem to have confirmed
important predictions of the Debye model, such as the temperature dependence $U_e(T) \propto T^{-1/2}$ \cite{HPR05}. But there are still controversies on the validity of the experimental analysis and the use of the Debye screening, or Salpeter formula, to describe the experiments \cite{RS95,Rol01}.  

A good measure of the screening effect is given by the screening parameter given  by $\Gamma = Z_1Z_2e^2/\langle r \rangle kT$, where $\langle r \rangle =n^{-1/3}$.  In the core of the sun densities are of the order of $\rho \sim 150$ g/cm$^3$ with temperatures of $T\sim 1.5 \times 10^7$ K.  For pp reactions in the sun, we thus get $\Gamma \sim 1.06$ which validates the weak screening approximation, but for p$^7$Be reactions one gets $\Gamma_{{\rm p}^7{\rm Be}}\sim 1.5$, which is one of the reasons to support modifications of the Salpeter formula. Also, in the sun the number of ions within a sphere of radius $R_D$ (Debye sphere) is of the order of $N\sim 4$. As the Debye-Hueckel approximation is based on the mean field approximation, i.e., for $N=n(4\pi R^3/3) \gg 1$, deviations from the Salpeter approximation are justifiable. 

\begin{figure}[t]
\begin{center}
{\includegraphics[width=9cm]{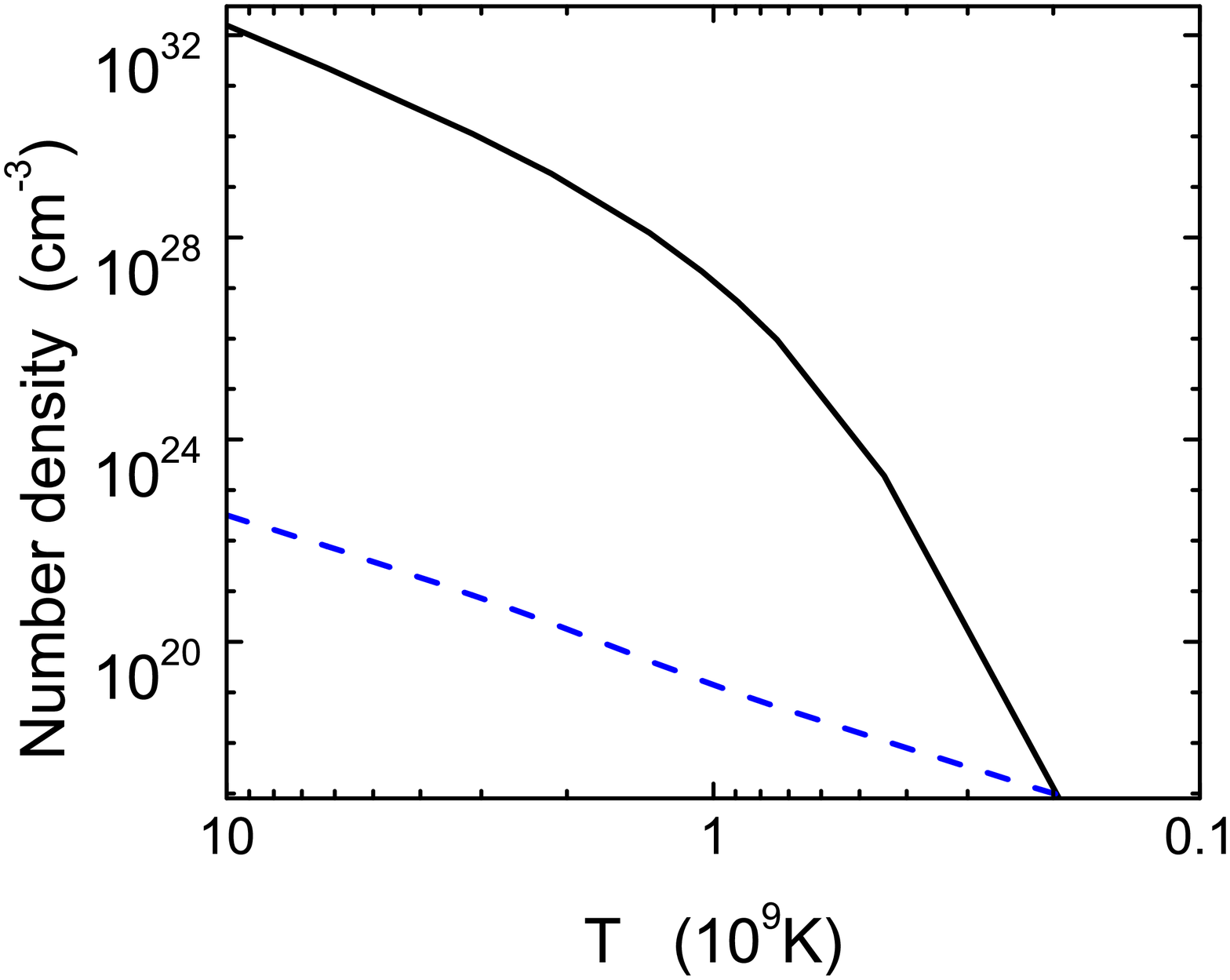}}
\end{center}
\vspace{-0.5cm}
\caption{
(Color online) Number density of electrons (solid curve) and of baryons (dashed curve) during the early universe as a function of the temperature in units of billion degrees Kelvin, $T_9$. }\label{fig1}
\end{figure}

Based on the discussion above, there is a possibility that the screening enhancement factor, eq. \eqref{f0}, could appreciably differ from the Salpeter formula under several circumstances, leading to non-negligible changes in the BBN and stellar evolution predictions. It is the purpose of this work to verify under what conditions would this statement be true. 

In this work, the BBN abundances were calculated with a modified version of the
standard BBN code derived from Wagoner, Fowler, and Hoyle \cite{WFH67}  and Kawano \cite{Kaw88,Kaw92} (for a public code, see \cite{Pisanti:2007hk}). 

The electron density during the early universe varies strongly with the temperature as seen in figure \ref{fig1}, where $T_9$ is the temperature in units of $10^9$ K ($T_9$). This can compared with the electron number density at the center of the sun, $n_e^{sun}\sim 10^{26}$/cm$^3$.  The figure shows that, at typical temperatures $T_9 \sim 0.1-1$ during the BBN the universe had electron densities which are much larger that the electron density in the sun.  However, in contrast to the sun, the baryon density in the early universe is much smaller than the electron density. The large electron density  is due to the $e^+e^-$ production by the abundant photons during the BBN. 

\begin{figure}[t]
\begin{center}
{\includegraphics[width=9cm]{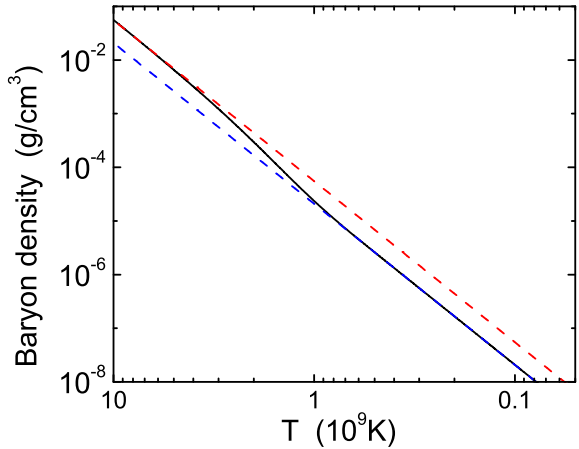}}
\end{center}
\vspace{-0.5cm}
\caption{
(Color online) Baryon density (solid curve) during the early universe as a function of the temperature in units of billion degrees Kelvin, $T_9$.   The dashed curves are obtained from Eq. \eqref{rhoT} with $h\sim 2.1\times 10^{-5}$ and $h\sim 5.7\times 10^{-5}$, respectively.  
}\label{fig2}
\end{figure}

The baryonic density is best seen in
figure \ref{fig2}. It varies as
\begin{equation}
\rho_b \simeq h T_9^3, 
\label{rhoT}
\end{equation} 
where $h$ is the baryon density parameter \cite{SW77}. It can be calculated  by using Eq. (3.11) of Ref. \cite{SW77} and the baryon-to-photon ratio $\eta=6.19\times 10^{-10}$  at the BBN epoch (from WMAP data \cite{Kom10}). Around $T_9\sim 2$ there is a change of the value of $h$ from $h\sim 2.1\times 10^{-5}$ to $h\sim 5.7\times 10^{-5}$.  Eq. \eqref{rhoT} with the two values of $h$ are shown as dashed lines in figure \ref{fig2}.

In Ref. \cite{Itoh97} a theory was developed to show how the abundant $e^+e^-$-pairs during the BBN, for temperatures below the neutrino decoupling ($T_9\sim 0.7$ MeV), result in modifications in the Salpeter formula. In fact, only a very small fraction of the electrons present in the medium is needed to neutralize the charge of the protons. The major part of the electrons are accompanied by the respective positrons created via $\gamma \gamma \rightarrow e^+e^-$ so that the total charge of the universe is zero.  At the decoupling temperature the neutron density is only about 1/6 of the total baryon density. Thus, the ion charge density (proton density) at this epoch is approximately equal to the baryon density. With this assumption the enhancement factor in Eq. \eqref{f02} becomes independent of the temperature, 
\begin{eqnarray}
f^{BBN}&=&\exp \left( 4.49 \times 10^{-8} \zeta Z_1 Z_2 \right)\nonumber \\
&\sim& 1+4.49 \times 10^{-8} \zeta Z_1 Z_2  \ \ \ {\rm for} \ \ T_9\lesssim  1,
\label{fbbn}
\end{eqnarray}
and
\begin{eqnarray}
f^{BBN}&=&\exp \left( 2.71 \times 10^{-8} \zeta Z_1 Z_2 \right)\nonumber \\
&\sim& 1+2.71 \times 10^{-8} \zeta Z_1 Z_2  \ \ \ {\rm for} \ \ T_9\gtrsim  2,
\end{eqnarray}
which yield small screening corrections for all known nuclear reactions. 

\begin{figure}[t]
\begin{center}
{\includegraphics[width=9cm]{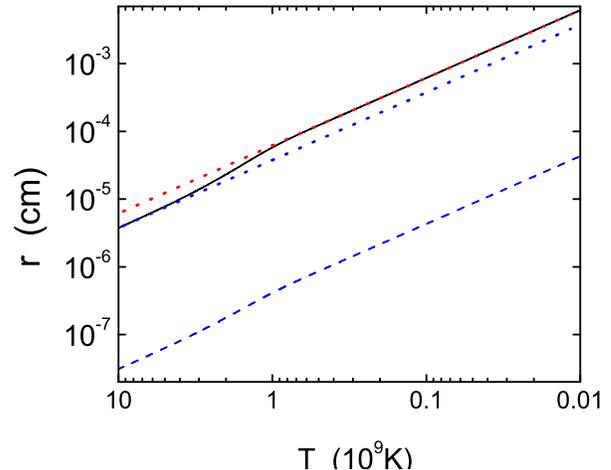}}
\end{center}
\vspace{-0.5cm}
\caption{
(Color online) Debye radius during the BBN as a function of the temperature in units of billion degrees Kelvin (solid line). The dotted lines are the approximation given by Eq. \eqref{rd} with  $R_D^{(0)}\sim 6.1\times 10^{-5}$ cm and $R_D^{(0)}\sim 3.7\times 10^{-5}$ cm, respectively. The inter-ion distance is shown by the isolated dashed-line. 
}\label{fig3}
\end{figure}

It is also worthwhile to calculate the Debye radius as a function of the temperature. This is shown in figure  \ref{fig3} where we plot Eq. \eqref{DR} with the ion density equal to  the proton density. The accompanying dashed lines correspond to the approximation of Eq. \eqref{rhoT}, with $h\sim 2.1\times 10^{-5}$ and $h\sim 5.7\times 10^{-5}$. This leads to two straight lines in a logarithmic plot of 
\begin{equation}
R_D = {R_D^{(0)}  T_9^{-1}},
\label{rd}
\end{equation}
with $R_D^{(0)}\sim 6.1\times 10^{-5}$ cm and $R_D^{(0)}\sim 3.7\times 10^{-5}$ cm, respectively. In figure \ref{fig3} we also show the inter-ion distance by the lower dashed line. It is clear that the number of ions inside the Debye sphere is  at least of the order $10^3$, which would justify the mean field approximation for the ions.
In contrast to protons, electrons and positrons are mostly relativistic and their chaotic motion will probably average out the effect of screening around the ions. But because the number density of electrons is large, an appreciable fraction of them still carry velocities comparable to those of the ions. The effects of such electrons on the modification of the Debye-Hueckel scenario might be worthwhile to investigate theoretically.

Based on the above arguments, the screening by electrons during the BBN is likely  a negligible effect. But one needs to verify arguments that sensitive quantities such as the Li/H ratio predicted by BBN might be impacted. This ratio is very small but is one of the major problems for the Big Bang predictions. In fact, there are discrepancies between the BBN theory and observation for the lithium isotopes, $^6$Li and $^7$Li. These discrepancies are substantiated by recent observations of metal-poor halo stars \cite{Asp06} and the high precision measurement of the baryon-to-photon ratio $\eta$ of the Universe by WMAP \cite{Kom10,Sper03,Ben03}. In view of the relevance of this topic to BBN and its predictions, it is worthwhile to check the influence of the electron screening on the elemental abundances.

The BBN is sensitive to certain parameters such as the baryon-to-photon ratio, number of neutrino families, and the neutron decay lifetime. We use the values $\eta=6.19 \times 10^{-10}$, $N_\nu=3$ and $\tau_n=878.5$ s for the baryon/photon ratio, number of neutrino families and neutron-day lifetime, respectively. We have included all reactions of the standard BBN model using the values of the cross sections as published in Ref. \cite{Nol}. The reaction rates were modified to include screening factors calculated with Eq. \eqref{fbbn}. 

In order to test under which circumstances the screening by electrons would make an appreciable impact on the BBN predictions, we have artificially modified the Slapeter formula for $f^{BBN}$ by rescaling it with a  fudge factor $w$, i.e.
\begin{equation}
f'=\;\mathrm{exp} \left(w{Z_1 \,Z_2 \;e^2 \over R_D k T}\right), \ \ \ \ \  {\rm or}  \ \ \ \ln f' =w\ln f,\label{w}
\end{equation}
with $w$ varying between 1 and $10^4$. We quantify the dependence of BBN on the electron screening by the quantity 
\begin{equation}
\Delta = {Y' -Y \over Y},
\end{equation}
where $Y$ denotes the abundance $Y$ of an element produced during the BBN.  The primed quantities $Y'$ denote the abundances modified by the screening effect. 

\begin{figure}[t]
\begin{center}
{\includegraphics[width=9cm]{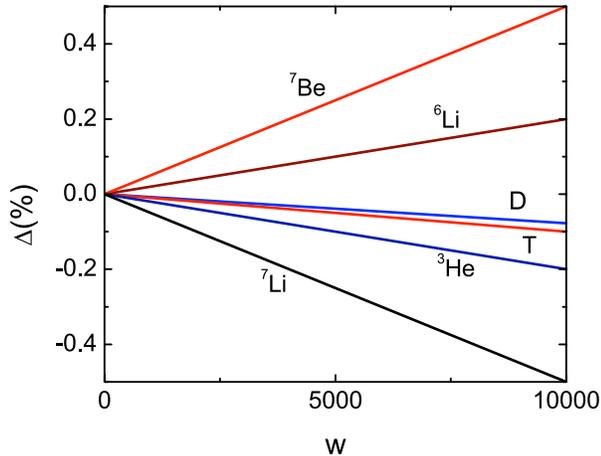}}
\end{center}
\vspace{-0.5cm}
\caption{
(Color online) Variation (in percent) of the abundances of several light nuclei  as  a function of a multiplicative factor $w$ artificially enhancing Salpeter's formulation of screening. $w$ is defined in Eq. \eqref{w}.   
}\label{fig4}
\end{figure}

In figure \ref{fig4} we show the value of $\Delta$ (in percent) as a function of the screening enhancement factor $w$ in Eq. \eqref{fbbn}. For $w\sim 1$ any of the abundances are modified by less than $10^{-5}$ in case that the screening factor is calculated according to Eq. \eqref{fbbn}. If the $w$ is increased the abundances of $^6$Li and $^7$Be increase, while those of D, T, $^3$He and $^7$Li decrease as $w$ increases. This shows that the elemental abundance ratios have a different sensitivity to the electron screening. These calculations also make it evident that the relative abundances in the BBN would be somewhat modified only if the electron screening would be enormously enhanced (by a factor $w$ larger than $10^4$) compared to the prediction of Salpeter's model.

In conclusion, using a standard numerical computation of the BBN we have shown that electron screening cannot be a source of measurable changes in the elemental abundance. This is verified by artificially increasing the screening obtained by traditional models \cite{Sal54}. We back our numerical results with very simple and transparent estimates. This is also substantiated by the mean-field calculations of screening due to the more abundant free $e^+e^-$ pairs published in Ref. \cite{Itoh97}. They conclude that screening due to free pairs might yield a 0.1\% change on the BBN abundances.   But even if mean field models for electron screening were not reliable under certain conditions, which we have discussed thoroughly in the text, it is extremely unlikely that electron screening might have any influence on the predictions of the standard Big Bang nucleosynthesis.

\medskip This work was partially supported by the U.S. Department of Energy grants
DE-FG02-08ER41533, DE-FC02-07ER41457 (UNEDF, SciDAC-2) and
the U.S. National Science Foundation Grant No. PHY-0855082.

\end{document}